\documentclass{fundam}
\usepackage[T1]{fontenc}
\usepackage{amssymb,mathtools,microtype,url,graphicx,tikz}
\usetikzlibrary{arrows.meta,calc}
\numberwithin{equation}{section}
\title{Edge-structure fault diameters of exchanged hypercubes}
\author{Lihui Dong\\
	Xi’an Precision Machinery Research Institute Kunming Branch \\ Kunming, China \\
	dlh1997999{@}163.com
	\and Zhuo Li\thanks{Corresponding author}\\
	Institute of Scientific and Technical Information of China \\
	Beijing, China \\
	liz{@}istic.ac.cn
	} 
\date{}
\begin{document}
\maketitle \runninghead{L. Dong, Z. Li}{Edge-structure fault diameters of exchanged hypercubes}
\begin{abstract}
We determine the edge-structure and edge-substructure fault diameters of the exchanged hypercube $EH(s,t)$ for every $1\le s\le t$. Both parameters equal $s+t+4$, two more than the fault-free diameter, except for $EH(2,2)$, where both are nine, and $EH(1,1)$, where they are five and six, respectively. This holds even though an admissible family may remove up to $2s$ vertices, more than the vertex connectivity $s+1$. For every $s\ge2$, the maximum is attained by $s$ parallel fault edges in class $0$. For $s\ge3$, the structure-fault diameter exceeds the classical $s$-vertex-fault diameter by exactly one.
\end{abstract}

\begin{keywords}
exchanged hypercube; structure fault diameter; substructure fault diameter; matching failures; fault-tolerant routing
\end{keywords}
\noindent\textbf{ACM Computing Classification:} Mathematics of computing: Graph theory.

\section{Introduction}

The exchanged hypercube, introduced by Loh, Hsu, and Pan~\cite{Loh2005}, retains short communication paths with fewer links than a hypercube of the same order. Its vertices are triples $(a,b,c)$, where $a$ is an $s$-bit string, $b$ is a $t$-bit string, and $c\in\{0,1\}$. An edge changes one $a$-coordinate when $c=0$, one $b$-coordinate when $c=1$, or the coordinate $c$ itself. We assume $1\le s\le t$ throughout. The vertices in classes $0$ and $1$ have degrees $s+1$ and $t+1$, respectively. Counting these degrees gives $2^{s+t+1}$ vertices and $(s+t+2)2^{s+t-1}$ edges. The hypercube of the same order has $(s+t+1)2^{s+t}$ edges, asymptotically twice as many, while its diameter is only one smaller: $D(EH(s,t))=s+t+2$~\cite{Tsai2014} and $D(Q_{s+t+1})=s+t+1$. This trade-off raises the question of how much the distances grow when vertices fail.

The classical fault diameter measures the largest residual distance after a prescribed number of vertex failures~\cite{Krishnamoorthy1987}. For an integer $q$ smaller than the vertex connectivity of $G$, write
\begin{equation}
 D_q^{\mathrm{vertex}}(G)
 =\max_{\substack{X\subseteq V(G)\\|X|\le q}}
   \operatorname{diam}(G-X).
 \label{eq:vertex-fault-definition}
\end{equation}
The connectivity of $EH(s,t)$ is $s+1$~\cite{Ma2010}. Further work has determined its super connectivity~\cite{MaZhu2011} and generalized connectivity under minimum-degree constraints on the remaining graph~\cite{LiXu2013}. Tsai, Chen, and Tan~\cite{Tsai2014} proved that the $s$-vertex-fault diameter and the $(s+1)$-wide diameter both equal $s+t+3$ for $3\le s\le t$.

Structure connectivity and substructure connectivity, introduced by Lin, Zhang, Fan, and Wang~\cite{Lin2016}, model failures occurring in connected groups of vertices. Zhao, Sabir, and Lin~\cite{Zhao2026} introduced the corresponding diameter parameters and determined them for subcube structures in hypercubes. In particular, for $n\ge4$, the edge-structure and edge-substructure fault diameters of $Q_n$ are both $n+1$~\cite[Theorem~3.6]{Zhao2026}. Zhao and Sabir~\cite{ZhaoSabir2026} also treated several star structures in hypercubes. The present paper determines these two parameters for exchanged hypercubes.

We use the vertex-disjoint fault convention of the structure-fault diameter model~\cite{Zhao2026,ZhaoSabir2026}. A $K_2$-structure family $\mathcal F$ consists of pairwise vertex-disjoint edges. A $K_2$-substructure family consists of pairwise vertex-disjoint copies of $K_1$ or $K_2$, the connected nonempty subgraphs of an edge. Its deletion removes the vertex set
\begin{equation}
 F_V=\bigcup_{H\in\mathcal F}V(H).
 \label{eq:removed-vertices}
\end{equation}
The structure connectivity $\kappa(G;K_2)$ is the minimum size of a $K_2$-structure family for which $G-F_V$ is disconnected or has at most one vertex. The substructure connectivity $\kappa^{\mathrm{sub}}(G;K_2)$ is defined by the same minimum over $K_2$-substructure families, with the superscript $\mathrm{sub}$ distinguishing this parameter from the dimension $s$. The associated fault diameters are
\begin{align}
 D_f(G;K_2)
 &=\max_{\substack{\mathcal F\text{ a }K_2\text{-structure family}\\
 |\mathcal F|\le\kappa(G;K_2)-1}}
 \operatorname{diam}(G-F_V),\label{eq:definition-df}\\
 D_f^{\mathrm{sub}}(G;K_2)
 &=\max_{\substack{\mathcal F\text{ a }K_2\text{-substructure family}\\
 |\mathcal F|\le\kappa^{\mathrm{sub}}(G;K_2)-1}}
 \operatorname{diam}(G-F_V).\label{eq:definition-df-sub}
\end{align}

Liu and Cheng~\cite{Liu2024} determined structure and substructure connectivities of exchanged hypercubes for several path, cycle, and star families. In particular, their path result~\cite[Theorem~3.1]{Liu2024} gives
\begin{equation}
 \kappa(EH(s,t);K_2)
 =\kappa^{\mathrm{sub}}(EH(s,t);K_2)=s+1
 \qquad(3\le s\le t).
 \label{eq:known-k2-connectivity}
\end{equation}
Immanuel and Greeni~\cite[Theorems~4 and~5]{Immanuel2023} obtained these values for $s\ge2$ and $t\ge3$. Pan~\cite{Pan2026} extended the study of star connectivity to larger stars and to step-limited variants. Proposition~\ref{prop:clean-connectivity} extends \eqref{eq:known-k2-connectivity} to all $1\le s\le t$, so both diameter parameters allow $s$ fault structures.

\begin{theorem}[Main result]\label{thm:main-exact}
For every $1\le s\le t$,
\begin{equation}
 D_f(EH(s,t);K_2)=
 \begin{cases}
 5,&(s,t)=(1,1),\\
 9,&(s,t)=(2,2),\\
 s+t+4,&\text{otherwise},
 \end{cases}\label{eq:main-edge}
\end{equation}
and
\begin{equation}
 D_f^{\mathrm{sub}}(EH(s,t);K_2)=
 \begin{cases}
 6,&(s,t)=(1,1),\\
 9,&(s,t)=(2,2),\\
 s+t+4,&\text{otherwise}.
 \end{cases}\label{eq:main-sub}
\end{equation}
\end{theorem}

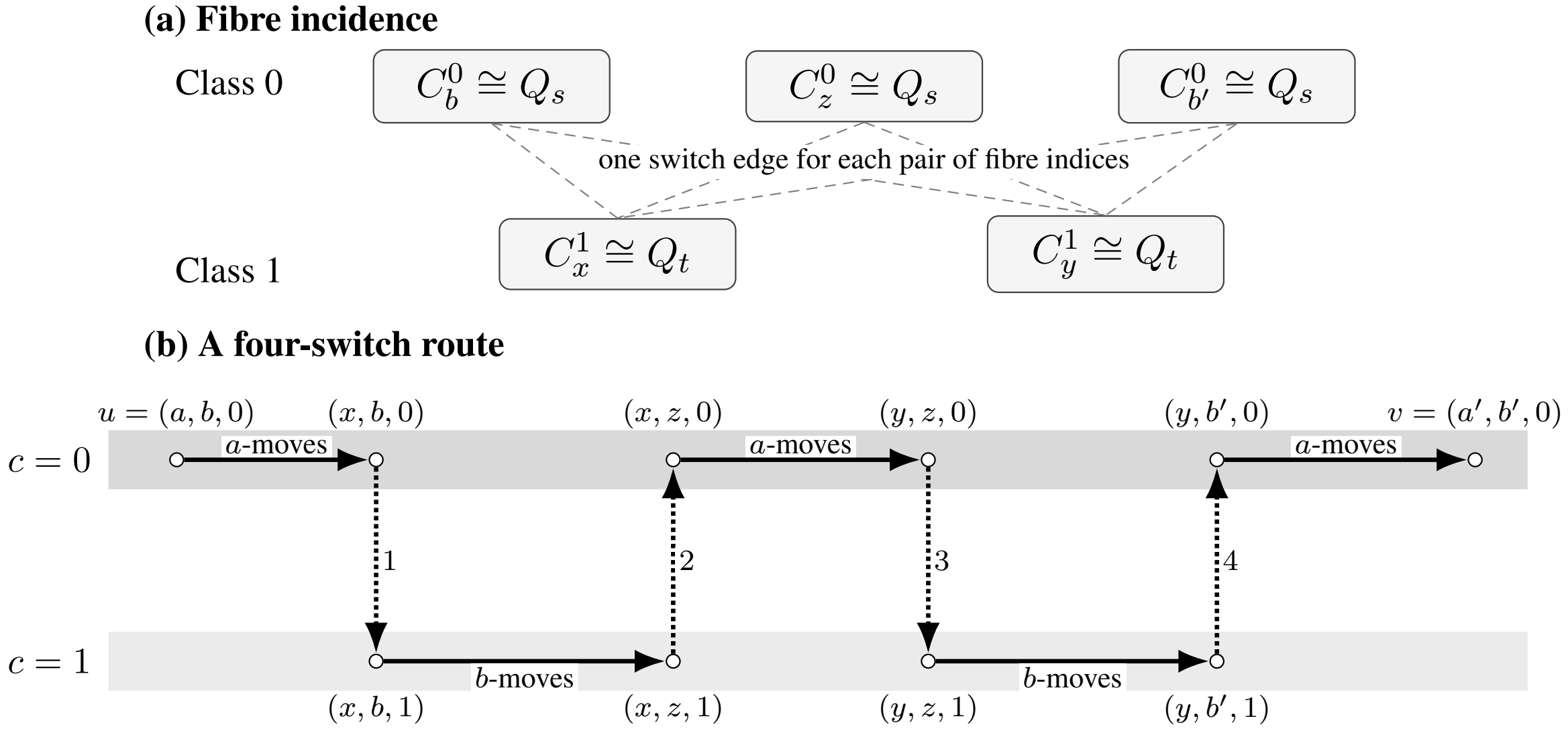
\begin{figure}[t]
\centering
\resizebox{\linewidth}{!}{%
\begin{tikzpicture}[x=1cm,y=1cm,>=Latex,
  fib/.style={draw=black!75,rounded corners=3pt,minimum width=2.25cm,minimum height=0.65cm,align=center,fill=black!4},
  route/.style={-Latex,very thick},
  cross/.style={-Latex,very thick,densely dotted},
  smalltag/.style={font=\scriptsize,fill=white,inner sep=1pt},
  vtx/.style={circle,draw=black,fill=white,inner sep=1.3pt}]
  \node[anchor=west,font=\bfseries\small] at (0,5.7) {(a) Fibre incidence};
  \node[font=\small] at (0.95,5.12) {Class 0};
  \node[font=\small] at (0.95,3.33) {Class 1};
  \node[fib] (b) at (3.45,5.08) {$C^0_b\cong Q_s$};
  \node[fib] (z) at (7.00,5.08) {$C^0_z\cong Q_s$};
  \node[fib] (bp) at (10.55,5.08) {$C^0_{b'}\cong Q_s$};
  \node[fib] (x) at (4.65,3.48) {$C^1_x\cong Q_t$};
  \node[fib] (y) at (9.30,3.48) {$C^1_y\cong Q_t$};
  \foreach \p in {b,z,bp}{\foreach \q in {x,y}{\draw[gray,densely dashed] (\p.south) -- (\q.north);}}
  \node[smalltag] at (7,4.36) {one switch edge for each pair of fibre indices};
  \node[anchor=west,font=\bfseries\small] at (0,2.60) {(b) A four-switch route};
  \draw[black!15,line width=16pt] (-0.20,1.52)--(13.32,1.52);
  \draw[black!8,line width=16pt] (-0.20,-0.40)--(13.32,-0.40);
  \node[anchor=east,font=\small] at (-0.23,1.52) {$c=0$};
  \node[anchor=east,font=\small] at (-0.23,-0.40) {$c=1$};
  \node[vtx] (u) at (0.45,1.52) {};
  \node[vtx] (p1) at (2.35,1.52) {};
  \node[vtx] (p2) at (5.18,1.52) {};
  \node[vtx] (p3) at (7.61,1.52) {};
  \node[vtx] (p4) at (10.36,1.52) {};
  \node[vtx] (v) at (12.82,1.52) {};
  \node[vtx] (q1) at (2.35,-0.40) {};
  \node[vtx] (q2) at (5.18,-0.40) {};
  \node[vtx] (q3) at (7.61,-0.40) {};
  \node[vtx] (q4) at (10.36,-0.40) {};
  \draw[route] (u)--node[above,smalltag] {$a$-moves} (p1);
  \draw[route] (q1)--node[below,smalltag] {$b$-moves} (q2);
  \draw[route] (p2)--node[above,smalltag] {$a$-moves} (p3);
  \draw[route] (q3)--node[below,smalltag] {$b$-moves} (q4);
  \draw[route] (p4)--node[above,smalltag] {$a$-moves} (v);
  \draw[cross] (p1)--node[right,smalltag] {$1$} (q1);
  \draw[cross] (q2)--node[right,smalltag] {$2$} (p2);
  \draw[cross] (p3)--node[right,smalltag] {$3$} (q3);
  \draw[cross] (q4)--node[right,smalltag] {$4$} (p4);
  \node[font=\scriptsize,above=0.16cm] at (u) {$u=(a,b,0)$};
  \node[font=\scriptsize,above=0.16cm] at (p1) {$(x,b,0)$};
  \node[font=\scriptsize,above=0.16cm] at (p2) {$(x,z,0)$};
  \node[font=\scriptsize,above=0.16cm] at (p3) {$(y,z,0)$};
  \node[font=\scriptsize,above=0.16cm] at (p4) {$(y,b',0)$};
  \node[font=\scriptsize,above=0.16cm] at (v) {$v=(a',b',0)$};
  \node[font=\scriptsize,below=0.18cm] at (q1) {$(x,b,1)$};
  \node[font=\scriptsize,below=0.18cm] at (q2) {$(x,z,1)$};
  \node[font=\scriptsize,below=0.18cm] at (q3) {$(y,z,1)$};
  \node[font=\scriptsize,below=0.18cm] at (q4) {$(y,b',1)$};
\end{tikzpicture}%
}
\caption{Fibre decomposition and a schematic four-switch route. (a) Each class-$0$ fibre $C^0_b$ is a copy of $Q_s$ and each class-$1$ fibre $C^1_a$ is a copy of $Q_t$; a unique switch edge joins $C^0_b$ to $C^1_a$, with endpoints $(a,b,0)$ and $(a,b,1)$. Each dashed line stands for one such edge. (b) A typical four-switch walk between two class-$0$ vertices. Horizontal arrows indicate paths within fibres, and dotted vertical arrows are switch edges.}
\label{fig:fibre-four-switch}
\end{figure}

In $EH(s,t)$, the $a$-coordinates change only in class $0$ and the $b$-coordinates only in class $1$. A route between two vertices must therefore alternate between the classes, and the fibre containing an endpoint can be cut off from the rest of its class even when the residual graph is connected. The upper-bound proof routes around such fibres. Every surviving vertex reaches a clean fibre of the other class in at most two steps (Lemma~\ref{lem:clean-fibre-access}), and the entrances at the two endpoints can be chosen so that their total coordinate cost is small (Lemma~\ref{lem:compatible-entrances}). Figure~\ref{fig:fibre-four-switch} shows the resulting four-switch route. When every cheap entrance is blocked, the blocking faults form an independent set that uses the whole budget, and this fixes their positions closely enough to give a different short route.

For the lower bound, $s$ parallel fault edges in class $0$ force every path between two suitable vertices to change class at least four times; this is the source of the term $+4$. The cases $s=1$ and $s=2$ are settled separately, and $EH(2,2)$ requires its own construction.

Section~\ref{sec:clean-fibre-upper} develops the clean-fibre argument. Section~\ref{sec:sharp-routing} proves the upper bound for $s\ge3$, and Section~\ref{sec:lower} gives the general matching construction. The remaining sections settle the smaller coordinate blocks and assemble the full classification.

\section{Clean fibres and connectivity}
\label{sec:clean-fibre-upper}

Write a vertex of $G=EH(s,t)$ as $(a,b,c)$, where $a\in\{0,1\}^{s}$, $b\in\{0,1\}^{t}$, and $c\in\{0,1\}$. At $c=0$, an edge changes one coordinate of $a$; at $c=1$, an edge changes one coordinate of $b$.  The remaining edges join $(a,b,0)$ to $(a,b,1)$.  Throughout this section, $1\le s\le t$. For $b\in\{0,1\}^{t}$ and $a\in\{0,1\}^{s}$, respectively, put
\begin{equation}
 C^0_b=G[\{(x,b,0):x\in\{0,1\}^{s}\}],
 \qquad
 C^1_a=G[\{(a,y,1):y\in\{0,1\}^{t}\}].
 \label{eq:clean-fibres}
\end{equation}
Thus $C^0_b\cong Q_s$ and $C^1_a\cong Q_t$.

Let $\mathcal F$ be a family of pairwise vertex-disjoint subgraphs, each isomorphic to $K_1$ or $K_2$, and let
\begin{equation}
 F_V=\bigcup_{H\in\mathcal F}V(H),\qquad q=|\mathcal F|\le s.
 \label{eq:clean-fault-family}
\end{equation}
A fibre is called clean if it contains no vertex of $F_V$; its index is then also called clean. The sets of indices of fibres that are not clean are
\begin{equation}
 \mathcal A=\{a:V(C^1_a)\cap F_V\ne\varnothing\},\qquad
 \mathcal B=\{b:V(C^0_b)\cap F_V\ne\varnothing\}.
 \label{eq:dirty-index-sets}
\end{equation}
Each member of $\mathcal F$ meets at most one class-$1$ fibre and at most one class-$0$ fibre.  Consequently,
\begin{equation}
 |\mathcal A|\le q\le s,\qquad |\mathcal B|\le q\le s\le t.
 \label{eq:dirty-index-count}
\end{equation}
We use $d_H$ for Hamming distance on binary strings of a fixed length. For a binary string $a$, its $i$th component is $a_i$, whereas $a^{(i)}=a\oplus e_i$ denotes the string obtained by changing that component. A switch edge changes only the class coordinate $c$. Every walk between two vertices contains a path between them of no greater length; upper bounds may therefore be proved by constructing walks.

\begin{lemma}[Access to a clean fibre]
\label{lem:clean-fibre-access}
Every surviving vertex $u=(a,b,0)$ has a path of length at most two in $G-F_V$ to a vertex $(a',b,1)$ in a clean fibre $C^1_{a'}$. Here $d_H(a,a')\le1$. Every surviving vertex $v=(a,b,1)$ has a path of length at most two to a vertex $(a,b',0)$ in a clean fibre $C^0_{b'}$, where $d_H(b,b')\le1$.
\end{lemma}

\begin{proof}
Consider $u=(a,b,0)$.  If $a\notin \mathcal A$, the edge from $u$ to $(a,b,1)$ gives the required path.  Suppose $a\in \mathcal A$. Let $e_i$ denote the $i$th unit vector of $\{0,1\}^{s}$, and put $a^{(i)}=a\oplus e_i$ and $u_i=(a^{(i)},b,0)$ for $1\le i\le s$. If $a^{(i)}\notin \mathcal A$ and $u_i\notin F_V$ for some $i$, then
\begin{equation}
 u,\ u_i,\ (a^{(i)},b,1)
 \label{eq:two-step-clean-access}
\end{equation}
is a surviving path to a clean fibre.

Suppose no such $i$ exists.  Since $a\in \mathcal A$, some faulty subgraph contains a class-$1$ vertex with first coordinate $a$. For each $i$, either $a^{(i)}\in \mathcal A$, witnessed by a faulty subgraph containing a class-$1$ vertex with first coordinate $a^{(i)}$, or $a^{(i)}\notin \mathcal A$ and a faulty subgraph contains $u_i$. These $s+1$ witnesses must belong to distinct faulty subgraphs. Indeed, a faulty subgraph cannot contain class-$1$ vertices from two different fibres.  It cannot contain both $u_i$ and $u_j$ for $i\ne j$, because these two vertices are nonadjacent.  Finally, a faulty subgraph containing both a class-$0$ and a class-$1$ vertex must be a switch edge, whose endpoints have the same first coordinate; it therefore cannot witness two different indices among $a,a^{(1)},\ldots,a^{(s)}$. This requires at least $s+1$ faulty subgraphs, contradicting $q\le s$.

The argument for a class-$1$ vertex exchanges the roles of the two coordinate blocks.  If all $t+1$ corresponding choices were blocked, they would require at least $t+1$ distinct faulty subgraphs, contradicting $q\le s\le t$.
\end{proof}

\begin{proposition}
\label{prop:clean-connectivity}
For every $1\le s\le t$, deleting at most $s$ pairwise vertex-disjoint $K_1$ or $K_2$ subgraphs leaves $EH(s,t)$ connected. Consequently,
\begin{equation}
 \kappa(EH(s,t);K_2)=\kappa^{\mathrm{sub}}(EH(s,t);K_2)=s+1.
 \label{eq:edge-structure-connectivity-all}
\end{equation}
\end{proposition}

\begin{proof}
Both clean index sets are nonempty: $2^s>s\ge|\mathcal A|$ and $2^t>t\ge|\mathcal B|$.  For every $a\notin \mathcal A$ and $b\notin \mathcal B$, the surviving switch edge
\begin{equation}
 (a,b,0)(a,b,1)
 \label{eq:clean-fibre-crossing}
\end{equation}
joins the clean fibres $C^0_b$ and $C^1_a$. Each clean fibre is connected.  Their union, together with these switch edges, is therefore connected.  By Lemma~\ref{lem:clean-fibre-access}, every surviving vertex reaches this connected subgraph. The residual graph has at least $2^{s+t+1}-2s>1$ vertices. This proves connectivity and the lower bound $s+1$ for both parameters in \eqref{eq:edge-structure-connectivity-all}.

For the upper bound, take $u=(0^s,0^t,0)$.  Let $e_i$ be the $i$th unit vector in $\{0,1\}^s$, and let $f_1$ be the first unit vector in $\{0,1\}^t$.  Delete the $s+1$ edges with endpoint sets
\begin{equation}
 \bigl\{(e_i,0^t,0),(e_i,0^t,1)\bigr\}
 \quad (1\le i\le s),
 \qquad
 \bigl\{(0^s,0^t,1),(0^s,f_1,1)\bigr\}.
 \label{eq:edge-isolation-cut}
\end{equation}
They are pairwise vertex-disjoint and delete every neighbour of $u$ while preserving $u$.  The distinct vertex $(0^s,1^t,0)$ also survives.  The residual graph is disconnected, proving the upper bound for the structure connectivity and hence also for the substructure connectivity.
\end{proof}

\begin{lemma}[A surviving intermediate coordinate]
\label{lem:good-hamming-intermediate}
Let $n\ge3$, let $Z\subseteq\{0,1\}^n$ satisfy $|Z|\le n$, and let $x,y\in\{0,1\}^n$.  There exists $z\notin Z$ such that
\begin{equation}
 d_H(x,z)+d_H(z,y)\le n.
 \label{eq:good-intermediate-bound}
\end{equation}
For $n=2$, the same assertion holds with $n+1$ on the right.
\end{lemma}

\begin{proof}
Put $h=d_H(x,y)$ and $\ell=n-h$.  The Hamming interval
\begin{equation}
 I(x,y)=\{z:d_H(x,z)+d_H(z,y)=h\}
 \label{eq:hamming-interval}
\end{equation}
has $2^h$ elements: the $h$ coordinates on which $x$ and $y$ differ are free, and the other coordinates are fixed. If $\ell\le1$ and $n\ge3$, then $|I(x,y)|\ge2^{n-1}>n\ge|Z|$, so a suitable $z$ lies in the interval.

If $\ell\ge2$, take the interval together with the strings obtained by changing exactly one of its $\ell$ fixed coordinates.  This set has $2^{h}(\ell+1)$ elements, and every element has distance sum at most $h+2\le n$.  Moreover,
\begin{equation}
 2^{h}(\ell+1)\ge(h+1)(\ell+1)>h+\ell=n.
 \label{eq:intermediate-candidate-count}
\end{equation}
It therefore contains an element outside $Z$.

For $n=2$, if $h=0$, the three strings at distance at most one from $x$ have distance sum at most two.  If $h=2$, every string has distance sum two.  If $h=1$, every string has distance sum at most three.  In all cases, more than two suitable strings are available.
\end{proof}

\section{Routing for coordinate blocks of dimension at least three}
\label{sec:sharp-routing}
Throughout this section, $3\le s\le t$, and at most $s$ pairwise vertex-disjoint copies of $K_1$ or $K_2$ are removed. Since a fault structure contains at most one vertex of any independent set, an independent set of removed vertices has size at most the number of fault structures. All upper-bound arguments below use this fact: if too many entrances are blocked, one exhibits more pairwise nonadjacent failed vertices, called \emph{witnesses}, than there are fault structures.

\subsection{Compatible entrances}
\begin{lemma}[Compatible entrances to clean opposite-class fibres]
\label{lem:compatible-entrances}
Fix a class and place the coordinate block that varies in that class first. Let its dimension be $m\geq3$, and let the fault budget $q$ be at most $m$. Write two surviving vertices in this order as $u=(a,b,c)$ and $v=(a',b',c)$, with $a,a'\in\{0,1\}^m$. There are labels $x,y$ at Hamming distance at most one from $a,a'$, respectively, such that:
\begin{enumerate}
 \item the vertices $(x,b,c)$ and $(y,b',c)$ survive, so $u$ and $v$ reach them in at most one step;
 \item the opposite-class fibres indexed by $x$ and $y$ are clean;
 \item their combined first-block cost satisfies
 \begin{equation}
  d_H(a,x)+d_H(x,y)+d_H(y,a')\leq m+1.
 \end{equation}
\end{enumerate}
\end{lemma}

\begin{proof}
It suffices to write the argument for class $c=0$; the other class follows by exchanging the roles of the coordinate blocks and of classes $0$ and $1$. A neighbouring label $a\oplus e_i$ is eligible for $u$ if its $c=0$ vertex at $b$-label $b$ survives and its opposite $c=1$ fibre is clean. If an index is ineligible, choose a failed witness with that $a$-label: either the failed $c=0$ vertex itself, or a failed vertex from its nonclean $c=1$ fibre. Every edge changes at most one $a$-coordinate, and the $a$-labels of witnesses chosen for distinct indices differ in two coordinates. Hence such witnesses are pairwise nonadjacent.

If the opposite-class fibre at $a$ is nonclean and no neighbouring index is eligible, its own failed $c=1$ witness and one witness for each of the $m$ ineligible indices form an independent set of $m+1$ failed vertices. This contradicts $q\leq m$. Consequently, a clean opposite-class fibre can always be reached by at most one first-block step followed by a switch.

If at least one of the fibres indexed by $a,a'$ is clean, use its original label and choose any eligible label at the other endpoint. The combined first-block cost is at most $m+1$. We may therefore assume both original opposite-class fibres are nonclean. Let $I,J$ be the nonempty sets of eligible neighbouring indices for $u,v$, and write $p=d_H(a,a')$.

If $p=0$, some index is eligible for both endpoints. Otherwise, for each of the $m$ indices choose an ineligibility witness from whichever endpoint blocks that index. Their $a$-labels are the $m$ distinct neighbours of $a$, so these witnesses are pairwise nonadjacent. Adding a failed vertex of the nonclean fibre $C^1_a$ gives an independent set of size $m+1$. With a common index $i$, the entrance labels coincide and the combined first-block cost is two.

If $p=1$, let $r$ be the differing coordinate. There is an index $i\ne r$ eligible for both endpoints. To prove this, suppose otherwise and choose one ineligibility witness for each of the $m-1$ indices other than $r$. Its label belongs to $\{a\oplus e_i,a'\oplus e_i\}$. For distinct indices, these chosen labels differ in either two or three coordinates. They therefore yield pairwise nonadjacent witnesses. Adding one failed $c=1$ witness at each of the labels $a,a'$ gives an independent set of $m+1$ failed vertices. The resulting contradiction proves the claim. Using the common index gives combined first-block cost three.

Suppose now that $p\geq2$. Each of $I,J$ has at least two indices. Indeed, if at most one index were eligible for $u$, choose failed witnesses for $m-1$ ineligible neighbours and add one failed vertex from each of the nonclean fibres $C^1_a$ and $C^1_{a'}$. The latter labels are distinct and neither is a neighbour label used in the witness set, since $p\geq2$. All $m+1$ witnesses are independent, a contradiction. The proof for $J$ is identical.

Let $D$ be the set of coordinates where $a,a'$ differ. If $p=m$, choose distinct indices $i\in I$, $j\in J$, possible because both sets have at least two elements. The entrance labels then have distance $m-2$, and the total first-block cost is $m$. Assume $p\leq m-1$. If $I\cap J\ne\varnothing$, use the same index at both endpoints, giving cost $p+2\leq m+1$. If the sets are disjoint and one contains an index of $D$, choose such an index and any index in the other set. At least one of the two entrance steps is toward the other endpoint, and the total cost is at most $p+2\leq m+1$. Finally, if $I,J$ are disjoint and both avoid $D$, their union contains at least four agreement coordinates. Thus $p+4\leq m$, and any pair of entrance indices has total first-block cost $p+4$. This proves all cases.
\end{proof}

\subsection{Equal-class endpoints}

\begin{theorem}[Equal-class endpoints]
\label{thm:equal-class-plus-four}
Let $3\leq s\leq t$. After at most $s$ vertex-disjoint copies of $K_1$ or $K_2$ fail in $EH(s,t)$, any two surviving vertices in the same $c$-class have distance at most
\begin{equation}
 s+t+4.
\end{equation}
\end{theorem}

\begin{proof}
Within this proof, write the block that varies in the endpoint class first and the other block second, relabelling them when $c=1$. Their dimensions are $m$ and $n$, respectively. Thus $(m,n)=(s,t)$ or $(t,s)$, and the fault budget $q$ is at most $\min(m,n)$, with $m,n\geq3$. Write the endpoints as $u=(a,b,c)$ and $v=(a',b',c)$, where $a,a'\in\{0,1\}^m$ are their first-block labels. Opposite-class fibres are indexed by first-block labels. An eligible entrance at an endpoint is its own first-block label, if the opposite-class fibre there is clean, or a neighbouring first-block label whose endpoint-class vertex survives and whose opposite-class fibre is clean.

Lemma~\ref{lem:compatible-entrances} provides eligible labels $x,y$ with combined first-block cost
\begin{equation}
 L=d_H(a,x)+d_H(x,y)+d_H(y,a')\leq m+1.
\end{equation}
If a clean endpoint-class fibre has second-block label $z$, the four-switch route through that fibre and the clean opposite fibres at $x,y$ has length
\begin{equation}
 L+d_H(b,z)+d_H(z,b')+4.
\end{equation}
At most $q$ second-block labels are nonclean.

If $b=b'$, among $b$ and its $n$ neighbours at least one label $z$ is clean, since $q\leq n$. Therefore the displayed length is at most $m+1+2+4=m+7\leq m+n+4$. Henceforth assume $b\ne b'$. Lemma~\ref{lem:good-hamming-intermediate} gives a clean second-block label $z$ with $d_H(b,z)+d_H(z,b')\leq n$. It therefore suffices to choose entrances with $L\leq m$, except in one saturated case handled directly below.

Put $p=d_H(a,a')$ and let $D$ be their set of differing first-block coordinates. If both original opposite-class fibres are clean, use $x=a,y=a'$, giving $L=p\leq m$.

\emph{Exactly one original opposite-class fibre is nonclean.} Suppose the fibre at $a$ is clean and the fibre at $a'$ is nonclean. Use $x=a$. An eligible neighbouring entrance $y=a'\oplus e_i$ exists by Lemma~\ref{lem:compatible-entrances}. Its cost is $p$ if $i\in D$, and $p+2$ if $i\notin D$. Thus $L\leq m$ unless $p=m-1$ and no index of $D$ is eligible. Let $j$ be the sole agreement coordinate in this exceptional case. The only eligible neighbouring index is then $j$.

Choose one failed witness in the nonclean opposite-class fibre at $a'$, and one failed witness for each of the $m-1$ ineligible labels $a'\oplus e_i$, $i\in D$. An ineligibility witness is either the failed endpoint-class vertex at second-block label $b'$, or any failed vertex in the nonclean opposite-class fibre at that first-block label. These $m$ witnesses are pairwise nonadjacent, and all their first-block labels have the original value of coordinate $j$. The independent witnesses imply $m\le q\le s\le m$, and hence $q=m=s$. In particular, this case cannot occur when $c=1$ and $s<t$. Every failed structure therefore contains exactly one of the witnesses. A failed structure lying wholly in the endpoint class must consequently lie in the endpoint-class fibre at $b'$. Outside that endpoint-class fibre, any failed endpoint-class vertex must belong to a switch structure and have the original value in coordinate $j$.

Use the eligible entrance $a'\oplus e_j$ at $v$, switch to its clean opposite-class fibre, and change the second-block label from $b'$ to $b$. Switch to the endpoint class. Since $b\ne b'$, every endpoint-class vertex in this fibre whose coordinate $j$ has been flipped survives. Change all $m-1$ differing first-block coordinates while keeping $j$ flipped, then flip $j$ back to arrive at the surviving vertex $u$. This path has length
\begin{equation}
 1+1+d_H(b',b)+1+(m-1)+1
 =m+d_H(b,b')+3\leq m+n+3.
\end{equation}
The case with the two endpoints interchanged is identical.

\emph{Both original opposite-class fibres are nonclean.} Let $I,J$ be the sets of eligible neighbouring indices. The independent-witness arguments in the proof of Lemma~\ref{lem:compatible-entrances} give the following facts: if $p=0$, there is an index eligible at both endpoints, giving $L=2$; if $p=1$, there is a common eligible index outside $D$, giving $L=3$; and if $p\geq2$, each of $I,J$ has at least two indices. The first two cases satisfy $L\leq m$ because $m\geq3$.

For $p=m$, choose distinct eligible indices at the two endpoints; both are in $D$ and give $L=m$. If $2\leq p\leq m-2$, an index in $I\cap J$ gives $L=p+2\leq m$. When the sets are disjoint and one contains an index in $D$, choose that index and any index in the other set; this also gives $L\leq p+2\leq m$. When the sets are disjoint and both avoid $D$, there are at least four agreement coordinates, and any choice gives $L=p+4\leq m$.

It remains that $p=m-1$. Let $j$ be the sole agreement coordinate. Since $|I|,|J|\geq2$, each set contains a differing coordinate. If distinct differing coordinates can be chosen at the two endpoints, they give $L=p\leq m$. Otherwise, both sets are exactly $\{i,j\}$ for one common differing coordinate $i$. Every index in $D\setminus\{i\}$ is then ineligible at both endpoints.

For each of these $m-2$ indices choose an ineligibility witness at each endpoint, and add one failed witness in each of the two original opposite-class fibres. The resulting $2m-2$ witnesses are pairwise nonadjacent. Within either endpoint's neighbour family, first-block labels differ in two coordinates. Endpoint-class witnesses from different families lie in distinct endpoint-class fibres because $b\ne b'$, so they are not adjacent. All first-block labels of the witnesses are distinct. Indeed, equality $a\oplus e_r=a'\oplus e_\ell$ would force $r\ne\ell$ and $D=\{r,\ell\}$. Since $r,\ell\in D\setminus\{i\}$, this contradicts $i\in D$. The original labels $a,a'$ are distinct from all these neighbour labels because $p=m-1\ge2$. Thus opposite-class witnesses are mutually nonadjacent, and a switch adjacency between the two classes is also impossible. This gives an independent set of $2m-2>m\geq q$ failed vertices, a contradiction.

All remaining cases therefore admit $L\leq m$, and the four-switch route has length at most $m+n+4=s+t+4$.
\end{proof}

\subsection{Opposite-class endpoints}

\begin{theorem}[Opposite-class endpoints]\label{thm:opposite-class-plus-four}
Let $3\le s\le t$, and let $\mathcal F$ be a family of at most $s$ pairwise vertex-disjoint copies of $K_1$ or $K_2$ in $EH(s,t)$. Write $F_V=\bigcup_{H\in\mathcal F}V(H)$. Every two surviving vertices in different coordinate classes satisfy
\begin{equation}
 d_{EH(s,t)-F_V}(u,v)\le s+t+4.
 \label{eq:opposite-class-plus-four}
\end{equation}
\end{theorem}

\begin{proof}
Write $u=(a,b,0)$ and $v=(a',b',1)$, and put
\begin{equation}
 p=d_H(a,a'),\qquad r=d_H(b,b').
 \label{eq:opposite-coordinate-distances}
\end{equation}
Call an $a$-label $\alpha$ an eligible entrance from $u$ if either $\alpha=a$ and $C^1_a$ is clean, or $\alpha$ is a Hamming neighbour of $a$, the vertex $(\alpha,b,0)$ survives, and $C^1_\alpha$ is clean. Define eligible $b$-labels $\beta$ from $v$ symmetrically: either $\beta=b'$ and $C^0_{b'}$ is clean, or $\beta$ is a neighbour of $b'$, the vertex $(a',\beta,1)$ survives, and $C^0_\beta$ is clean. Lemma~\ref{lem:clean-fibre-access} guarantees an eligible label at each endpoint.

For eligible $\alpha,\beta$, take the entrance step from $u$, switch to the clean fibre $C^1_\alpha$, change the $b$-label to $\beta$, switch to the clean fibre $C^0_\beta$, change the $a$-label to $a'$, switch to $(a',\beta,1)$, and reverse the optional $b$-coordinate entrance step from $v$. This gives a surviving walk of length
\begin{equation}
 3+H(\alpha)+L(\beta),\qquad
 H(\alpha)=d_H(a,\alpha)+d_H(\alpha,a'),\quad
 L(\beta)=d_H(b,\beta)+d_H(\beta,b').
 \label{eq:opposite-three-switch-route}
\end{equation}

For the $a$-coordinate cost, an eligible original label or an eligible neighbour obtained by flipping a coordinate on which $a,a'$ differ gives $H=p$. A flip in an agreement coordinate gives $H=p+2$. Consequently some eligible entrance always satisfies $H\le s+1$. Moreover, $H\le s$ unless $p=s-1$ and every label in
\begin{equation}
 \mathcal U=\{a\}\cup\{a\oplus e_i:a_i\ne a'_i\}
 \label{eq:high-cheap-entrances}
\end{equation}
is ineligible. Indeed, if $p=s$, every neighbouring entrance flips a coordinate on which $a,a'$ differ, while if $p\le s-2$, even a flip of an agreement coordinate gives $H\le s$.

The $b$-coordinate argument gives $L\le t+1$, with $L\le t$ unless $r=t-1$ and every label in
\begin{equation}
 \mathcal V=\{b'\}\cup\{b'\oplus f_j:b_j\ne b'_j\}
 \label{eq:low-cheap-entrances}
\end{equation}
is ineligible. Here $e_i,f_j$ are unit vectors in the two respective coordinate blocks.

At a single endpoint, each faulty structure blocks at most one of the original label and its neighbouring labels. Indeed, a label is blocked either because its opposite-class fibre is nonclean or because its endpoint-class vertex is deleted. A structure meets at most one opposite-class fibre; it cannot delete two neighbours of the endpoint, since these are nonadjacent; and a switch edge keeps both coordinate labels, so it cannot delete one entrance vertex while meeting the opposite-class fibre of another. Hence the number of blocked labels at an endpoint is at most the number of faulty structures.

Suppose first that $t>s$. The bound $L\le t$ can fail only if all $t$ labels in $\mathcal V$ are blocked, which needs at least $t>s$ structures. Therefore $L\le t$, and \eqref{eq:opposite-three-switch-route} has length at most $3+(s+1)+t=s+t+4$.

It remains to consider $s=t=m$. If $H\le m$ or $L\le m$ for some eligible choice, the same estimate proves the result. Otherwise,
\begin{equation}
 p=r=m-1,
 \qquad |\mathcal U|=|\mathcal V|=m,
 \label{eq:double-saturated-entrances}
\end{equation}
and all choices in both sets are blocked. The counting property forces $|\mathcal F|=m$, and every faulty structure must block one choice from $\mathcal U$ and one choice from $\mathcal V$. Since $m\ge3$, the endpoint distances $m-1$ are at least two, so
\begin{equation}
 a'\notin \mathcal U,\qquad b\notin \mathcal V.
 \label{eq:outside-cheap-entrance-sets}
\end{equation}

A fault contained entirely in class $0$ can block an $a$-entrance only by deleting one of the class-$0$ neighbours of $u$. It must therefore lie in the class-$0$ fibre $C^0_b$. To block a $b$-entrance, that same fault would have to meet a class-$0$ fibre whose index belongs to $\mathcal V$. This is impossible because $b\notin \mathcal V$. Symmetrically, a fault contained entirely in class $1$ would have to lie in $C^1_{a'}$ to block a $b$-entrance and meet a class-$1$ fibre indexed in $\mathcal U$ to block an $a$-entrance, contrary to $a'\notin \mathcal U$. This also excludes singleton faults. Hence every faulty structure is a switch edge, necessarily with $a$-label in $\mathcal U$ and $b$-label in $\mathcal V$.

It follows from \eqref{eq:outside-cheap-entrance-sets} that the entire class-$0$ fibre $C^0_b$ and the entire class-$1$ fibre $C^1_{a'}$ are clean. Travel from $u$ to $(a',b,0)$ within the former, cross to $(a',b,1)$, and travel to $v$ within the latter. This path has length
\begin{equation}
 p+1+r=2m-1\le s+t+4.
 \label{eq:double-saturated-short-route}
\end{equation}
The exceptional case is therefore covered as well.
\end{proof}

\begin{corollary}[Sharp general upper bound]\label{cor:sharp-upper}
For $3\le s\le t$, deleting at most $s$ pairwise vertex-disjoint $K_1$ or $K_2$ structures leaves a graph of diameter at most $s+t+4$.
\end{corollary}
\begin{proof}
Every pair of surviving vertices belongs either to the same coordinate class or to different coordinate classes. Apply Theorems~\ref{thm:equal-class-plus-four} and \ref{thm:opposite-class-plus-four}, respectively.
\end{proof}

\begin{remark}[Constructive routing from coordinate labels]
\label{rem:routing-algorithm}
Suppose $3\le s\le t$. Given the coordinate labels of the faulty structures and of two surviving vertices, a path of length at most $s+t+4$ can be constructed in $O((s+t)^3)$ bit operations and $O((s+t)^2)$ space.

Put $N=s+t$. The fault list and the sets of nonclean fibre labels have $O(N)$ entries, each of length $O(N)$. Direct comparisons test whether a vertex survives or a fibre is clean in $O(N^2)$ bit operations. The eligible entrances at each endpoint are found by at most $N+1$ such tests. There are at most $O(N^2)$ pairs of entrances; their coordinate costs can be computed in $O(N)$ bit operations per pair.

For equal-class endpoints, use the local block order and dimensions $m,n$ from Theorem~\ref{thm:equal-class-plus-four}, and minimize the combined first-block cost $L$. Use the four-switch route when $b=b'$ or $L\le m$. Otherwise $b\ne b'$ and $L=m+1$; use the two-switch route of the saturated case in that theorem. For opposite-class endpoints, minimize $H(\alpha)$ and $L(\beta)$ independently. Use the three-switch route when their sum is at most $s+t+1$, and otherwise use the one-switch route of the double-saturated case in Theorem~\ref{thm:opposite-class-plus-four}.

To choose a clean intermediate label between $b,b'\in\{0,1\}^n$, put $h=d_H(b,b')$ and $\ell=n-h$. If $\ell\le1$, enumerate only the words of the Hamming interval $I(b,b')$, using a binary counter on its free coordinates. If $\ell\ge2$, enumerate the interval words together with the words obtained by changing exactly one fixed coordinate. In either case, stop after $n+1$ distinct candidates. Lemma~\ref{lem:good-hamming-intermediate} shows that at least $n+1$ such candidates exist and that each has distance sum at most $n$. At most $n$ are forbidden, so direct membership tests find an admissible label in $O(N^3)$ bit operations. For the same-fibre case, only the second-block label and its neighbours need be checked.

The route vertices are generated by single-coordinate changes. Every resulting walk has $O(N)$ vertices. Repeated portions can be removed by storing the walk in a stack and comparing each new label against the current stack. This also takes $O(N^3)$ bit operations. The fault lists, entrance lists, and walk use $O(N^2)$ space.
\end{remark}
\section{A matching that forces four class changes}
\label{sec:lower}

\begin{theorem}\label{thm:general-lower}
For $2\le s\le t$, there is a matching of $s$ edges, all contained in class $0$, whose endpoint deletion leaves a pair of vertices at distance at least $s+t+4$.
\end{theorem}
\begin{proof}
Fix a coordinate $j\in\{1,\ldots,s\}$, and put
\begin{equation}
 u=(0^s,0^t,0),\qquad v=(1^s,1^t,0).
 \label{eq:lower-endpoints}
\end{equation}
Delete the $s-1$ edges
\begin{equation}
 \{(e_i,0^t,0),(e_i\oplus e_j,0^t,0)\},
 \qquad i\in\{1,\ldots,s\}\setminus\{j\},
 \label{eq:lower-initial-matching}
\end{equation}
and the edge
\begin{equation}
 \{(0^s,1^t,0),(e_j,1^t,0)\}.
 \label{eq:lower-terminal-edge}
\end{equation}
These $s$ edges are pairwise vertex-disjoint: the endpoints of the edge indexed by $i$ have $a$-labels with supports $\{i\}$ and $\{i,j\}$, which differ for different $i$, and the final edge lies in a different class-$0$ fibre. Both endpoints $u,v$ survive, since $s\ge2$.

Inside the initial class-$0$ fibre, the component of $u$ has $a$-labels exactly $\{0^s,e_j\}$. Every other neighbour of these two vertices was deleted by~\eqref{eq:lower-initial-matching}. Both labels are deleted in the terminal class-$0$ fibre by \eqref{eq:lower-terminal-edge}. Any path using exactly two changes of $c$ would have just one class-$1$ segment. That segment preserves its $a$-label, so it would have to leave the initial fibre with a label in $\{0^s,e_j\}$ and enter the terminal fibre at a deleted vertex. This is impossible. A path with no change of $c$ cannot change its $b$-label. Since both endpoints have $c=0$, the number of class changes is even and is therefore at least four.

The endpoint coordinate blocks differ in $s+t$ positions, requiring at least $s+t$ coordinate-changing edges in addition to the class changes. Every surviving path has length at least $s+t+4$.
\end{proof}

\begin{corollary}\label{cor:exact-large}
For every $3\le s\le t$,
\begin{equation}
 D_f(EH(s,t);K_2)=D_f^{\mathrm{sub}}(EH(s,t);K_2)=s+t+4.
 \label{eq:exact-large}
\end{equation}
The maximum is already attained by a matching contained entirely in class $0$.
\end{corollary}
\begin{proof}
The common failure budget is $s$ by Proposition~\ref{prop:clean-connectivity}. The matching in Theorem~\ref{thm:general-lower} is admissible for both fault models, and Corollary~\ref{cor:sharp-upper} supplies the common upper bound.
\end{proof}

\section{The one-dimensional coordinate block}

We use the following elementary hypercube observation.
\begin{lemma}\label{lem:cube-clique-deletion}
Let $t\ge2$. If $Z$ is empty, a singleton, or the two endpoints of an edge of $Q_t$, then $Q_t-Z$ is an isometric subgraph of $Q_t$: the distance between any two surviving vertices equals their Hamming distance.
\end{lemma}
\begin{proof}
Take surviving vertices at Hamming distance $h$. For $h\le1$ the assertion is immediate. For $h=2$, the two possible internal vertices of a shortest path are nonadjacent, so they cannot both belong to $Z$. For $h\ge3$, there are $h$ internally vertex-disjoint shortest paths: list the $h$ differing coordinates cyclically and, for each starting coordinate, flip them in that cyclic order. At each internal step the flipped coordinates form a cyclic interval of fixed length, and different starting coordinates give distinct intervals; vertices occurring at different steps have different Hamming distance from the initial vertex. At most two paths meet $Z$, so at least one shortest path survives.
\end{proof}

For $EH(1,t)$ write
\begin{equation}
 A_b=(0,b,1),\quad X_b=(0,b,0),\quad
 Y_b=(1,b,0),\quad B_b=(1,b,1).
\end{equation}
The $A$-vertices and the $B$-vertices induce two copies of $Q_t$. For each $b$, they are joined by a path $A_bX_bY_bB_b$ of length three. These paths are internally disjoint.

\begin{theorem}\label{thm:s1-exact}
For every $t\ge2$,
\begin{equation}
 D_f(EH(1,t);K_2)=D_f^{\mathrm{sub}}(EH(1,t);K_2)=t+5.
\end{equation}
For $EH(1,1)=C_8$, the corresponding values are $5$ and $6$, respectively.
\end{theorem}

\begin{proof}
Both connectivity parameters equal two, so it suffices to consider no fault, one vertex fault, and one edge-structure fault. By symmetry, edge faults have three forms: an edge inside the $A$-fibre, the edge $A_0X_0$, or the edge $X_0Y_0$.

First suppose that the removed edge lies inside the $A$-fibre. The $B$-fibre is clean, and every surviving vertex has a path to $B_b$ of length at most $3$, according to its type: $0$ for $B_b$, $1$ for $Y_b$, $2$ for $X_b$, and $3$ for $A_b$. For any pair that does not consist of two $A$-vertices, the two attachment lengths sum to at most $5$. A path through the $B$-fibre therefore has length at most $t+5$. For two surviving $A$-vertices, Lemma~\ref{lem:cube-clique-deletion} gives a path of length at most $t$ inside the $A$-fibre.

The same argument applies when the removed edge is $A_0X_0$: every surviving vertex still has the indicated attachment path to the $B$-fibre, and the surviving $A$-fibre is isometric after one vertex is removed. It also applies when the sole fault is the vertex $A_0$.

Next suppose that the removed edge is $X_0Y_0$. The preceding attachment argument applies to all surviving vertices except $A_0$. Pairs of $A$-vertices have distance at most $t$ in the clean $A$-fibre. For $b\ne0$, paths from $A_0$ through $A_b$ to $X_b$, $Y_b$, and $B_b$ have lengths $|b|+1$, $|b|+2$, and $|b|+3$, respectively. For $B_0$, use a unit vector $f$ and the path
\begin{equation}
 A_0,A_f,X_f,Y_f,B_f,B_0,
\end{equation}
of length five. All these bounds are at most $t+5$.

If the sole fault is $X_0$, the same argument applies except that $Y_0$ now survives. The path just displayed, followed by $B_0Y_0$, has length six, again at most $t+5$ because $t\ge2$. Symmetry covers single faults in the other two vertex types. The case of no fault also satisfies the bound: the fault-free graph has diameter $t+3$.

For the lower bound, delete the $A$-fibre edge $A_0A_f$, with $f$ a unit vector. Let $u=X_0$ and $v=A_{\mathbf1}$, where $\mathbf1=1^t$. Since $t\ge2$, the vertex $v$ survives. Every path from $u$ must begin with $X_0Y_0B_0$. It must then reach an $A$-vertex, and hence traverse a surviving connecting path of length three between the two fibres. At least $t$ edges that change $b$ are needed, because the endpoint $b$-labels are $0^t$ and $1^t$. The total length is therefore at least $2+3+t=t+5$. The path $X_0Y_0B_0$, followed by a shortest path in the clean $B$-fibre to $B_{\mathbf1}$ and then by $B_{\mathbf1}Y_{\mathbf1}X_{\mathbf1}A_{\mathbf1}$, attains this length. The same edge fault is admissible for the substructure parameter.

Finally, deleting one edge structure from $C_8$ leaves $P_6$, whose diameter is five, whereas deleting one vertex leaves $P_7$, whose diameter is six. This proves the exceptional values.
\end{proof}

\section{The two-dimensional coordinate block}

\begin{theorem}\label{thm:s2-exact}
For every integer $t\ge2$,
\begin{equation}
 D_f(EH(2,t);K_2)=D_f^{\mathrm{sub}}(EH(2,t);K_2)
 =\max\{t+6,9\}
 =\begin{cases}
 9,&t=2,\\
 t+6,&t\ge3.
 \end{cases}
\end{equation}
\end{theorem}

\begin{proof}
Both connectivity parameters equal three. We prove the upper bound for an arbitrary family of at most two pairwise vertex-disjoint $K_1$ or $K_2$ structures. The lower-bound families consist of edges alone. The class-$0$ fibres $C^0_b$ are copies of $Q_2$, and the class-$1$ fibres $C^1_a$ are copies of $Q_t$. Each structure meets at most one fibre of each class. Let $M_t=\max\{t+6,9\}$.

\paragraph{An oriented form of clean access.}
Take a surviving class-$1$ vertex $v=(a',b',1)$ and a target label $b\in\{0,1\}^t$. Put $h=d_H(b,b')$. If $C^0_{b'}$ is clean, set $\beta=b'$. If $C^0_{b'}$ is not clean and $h\ge2$, consider the $h$ coordinates on which $b$ and $b'$ differ. At least one such coordinate $j$ has both $(a',b'\oplus f_j,1)$ surviving and $C^0_{b'\oplus f_j}$ clean. Otherwise, as in the proof of Lemma~\ref{lem:clean-fibre-access}, the $h$ blocked labels and the nonclean fibre $C^0_{b'}$ would need $h+1\ge3$ distinct structures, exceeding the budget of two. For such a coordinate $j$, set $\beta=b'\oplus f_j$.

Thus, when $h\ge2$, $v$ reaches a clean $C^0_\beta$ by changing at most one $b$-coordinate and then changing the class, and
\begin{equation}
 d_H(b',\beta)+d_H(b,\beta)=h.
\end{equation}
When $h\le1$, ordinary clean access gives such a label with
\begin{equation}
 d_H(b',\beta)+d_H(b,\beta)\le h+2\le3.
\end{equation}
In all cases the indicated access path survives.

\paragraph{Upper bound: different classes.}
Let $u=(a,b,0)$ and $v=(a',b',1)$ survive. By Lemma~\ref{lem:clean-fibre-access}, $u$ reaches a clean class-$1$ fibre $C^1_\alpha$ by changing at most one $a$-coordinate and then changing the class. Choose the clean class-$0$ fibre $C^0_\beta$ for $v$ using the preceding oriented access argument. Route from $u$ to $C^1_\alpha$, along that class-$1$ fibre to the vertex $(\alpha,\beta,1)$, across to $C^0_\beta$, along that fibre to $(a',\beta,0)$, and then reverse the access path from $v$. There are three class changes. The total number of $a$-coordinate changes is at most
\begin{equation}
 d_H(a,\alpha)+d_H(\alpha,a')\le1+2=3.
\end{equation}
If $d_H(b,b')\ge2$, the total number of $b$-coordinate changes is $d_H(b,b')\le t$. Otherwise it is at most three. The resulting surviving walk has length at most $3+3+\max\{t,3\}=M_t$.

\paragraph{Upper bound: two class-$0$ vertices.}
Let $u=(a,b,0)$ and $v=(a',b',0)$ survive. If either endpoint class-$0$ fibre is clean, access a clean class-$1$ fibre from the other endpoint in at most two steps, follow that class-$1$ fibre to the clean class-$0$ fibre, change the class, and finish inside the class-$0$ fibre. This costs at most $2+t+1+2=t+5\le M_t$.

Suppose first that $b=b'$. If the two endpoints lie in the same component of the surviving $4$-cycle $C^0_b$, their distance within that fibre is at most two. Otherwise they are its opposite vertices, and the other two vertices have been removed by different structures. Neither of these structures can be an edge inside $C^0_b$, because each removed vertex has $u,v$ as its two neighbours in class $0$. Each structure is therefore a singleton or its switch edge. The class-$1$ fibres $C^1_a,C^1_{a'}$ are consequently clean, as is every class-$0$ fibre other than $C^0_b$. Choose a $b$-label $\beta$ adjacent to $b$. Routing through $C^1_a,C^0_\beta,C^1_{a'}$ uses four class changes, two $b$-coordinate changes, and at most two $a$-coordinate changes, for a total of at most eight.

Now assume that $b\ne b'$ and that both endpoint class-$0$ fibres are not clean. Exactly one fault structure meets each class-$0$ fibre, and these are all the faults. Each is a class-$0$ singleton, a class-$0$ edge, or a switch edge. Let $q_\times$ count the switch fault edges.

If $q_\times=0$, all class-$1$ fibres are clean and the only nonclean class-$0$ fibres are $C^0_b,C^0_{b'}$. There exists a clean $b$-label $\beta$ with
\begin{equation}
 d_H(b,\beta)+d_H(\beta,b')\le\max\{t,3\}.
\end{equation}
To see this, if $d_H(b,b')\ge2$, choose an internal label on a shortest $b$--$b'$ path; if $d_H(b,b')=1$, change from $b$ a coordinate different from the one on which $b,b'$ differ. This gives a label at distances one and two from the endpoints. The route through $C^1_a,C^0_\beta,C^1_{a'}$ now has length at most $4+2+\max\{t,3\}=M_t$.

If $q_\times=1$, orient the notation so that the fault in $C^0_{b'}$ is a switch edge with $a$-label $\gamma$, while the fault in $C^0_b$ has no class-$1$ vertex. The only nonclean class-$1$ fibre is $C^1_\gamma$. The surviving part of $C^0_b$ is connected and has at least two vertices. From $u$, choose in at most one $a$-coordinate step a surviving label $\alpha\ne \gamma$. If $a\ne \gamma$, simply take $\alpha=a$. Use the clean class-$1$ fibre $C^1_\alpha$ to reach $C^0_{b'}$ and finish inside that class-$0$ fibre, which has only its vertex with label $\gamma$ deleted. Its surviving subgraph is isometric, so the total length is at most $1+1+t+1+2=t+5$.

If $q_\times=2$, each endpoint class-$0$ fibre has lost one vertex and at most two class-$1$ fibres are nonclean. Choose a clean class-$1$ fibre $C^1_\alpha$. Both class-$0$ fibre subgraphs are isometric, and routing through $C^1_\alpha$ costs at most $2+1+t+1+2=t+6$. This covers all pairs of class-$0$ vertices.

\paragraph{Upper bound: two class-$1$ vertices.}
When $t=2$, exchanging the two coordinate blocks and complementing the class bit is an automorphism of the graph, so the preceding class-$0$ argument applies. Assume below that $t\ge3$ and let $u=(a,b,1)$ and $v=(a',b',1)$ survive.

If one endpoint class-$1$ fibre is clean, access a clean class-$0$ fibre from the other endpoint, follow that class-$0$ fibre to the clean class-$1$ fibre, change the class, and finish inside it. The length is at most $2+2+1+t=t+5$.

Suppose that $a=a'$. If at most one fault structure meets $C^1_a$, its deleted vertices form a singleton or the endpoints of one edge. Lemma~\ref{lem:cube-clique-deletion} gives the original Hamming distance, at most $t$, between $u,v$ inside this class-$1$ fibre. If both fault structures meet $C^1_a$, every other class-$1$ fibre is clean. Moreover, both endpoint class-$0$ fibres $C^0_b,C^0_{b'}$ are clean. A structure meeting $C^1_a$ can contain a class-$0$ vertex only if it is a switch edge; such an edge with $b$-label $b$ or $b'$ would also delete the endpoint $u$ or $v$, respectively. Choose an $a$-label $\alpha$ adjacent to $a$. Route from $u$ through the clean $C^0_b$, then through the clean $C^1_\alpha$, and finally through the clean $C^0_{b'}$ to $v$. The route uses four class changes, two $a$-coordinate changes, and $d_H(b,b')$ $b$-coordinate changes. Its length is at most $t+6$.

It remains to consider distinct nonclean class-$1$ fibres $C^1_a,C^1_{a'}$. Exactly one fault structure meets each class-$1$ fibre, and these are all the faults. Each structure is a class-$1$ singleton, a class-$1$ edge, or a switch edge. The surviving part of each endpoint class-$1$ fibre is isometric by Lemma~\ref{lem:cube-clique-deletion}. For a fault structure, its set of $b$-labels is a singleton or two adjacent vertices of $Q_t$. Such a set contains at most one neighbour of $b$, because the neighbours of a hypercube vertex are independent. The two structures together therefore cover at most two of the $t$ neighbours of $b$. Since $t\ge3$, choose a neighbour $\beta$ of $b$ whose label occurs in neither structure. Both endpoint class-$1$ fibres survive at $\beta$. The entire class-$0$ fibre $C^0_\beta$ is clean: every class-$0$ vertex among these faults comes from a switch fault, whose $b$-label has been excluded. Route within $C^1_a$ to $\beta$, across and through $C^0_\beta$, and back into $C^1_{a'}$ to reach $v$. The two isometric class-$1$ fibre segments have total length
\begin{equation}
 1+d_H(\beta,b')\le d_H(b,b')+2\le t+2.
\end{equation}
There are two class changes and at most two $a$-coordinate changes, giving a total at most $t+6$. The upper bound $M_t$ is now proved for every surviving pair.

\paragraph{Lower bound $t+6$.}
For any $t\ge2$, delete the two vertex-disjoint edges
\begin{equation}
 \{(00,0^t,1),(00,f_1,1)\},\qquad
 \{(10,0^t,0),(11,0^t,0)\}.
\end{equation}
Let $u=(00,0^t,0)$ and $v=(10,1^t,1)$. Every simple path from $u$ begins
\begin{equation}
 (00,0^t,0)\to(01,0^t,0)\to(01,0^t,1).
\end{equation}
The first class-$0$ vertex has only the indicated surviving neighbour; the next vertex has only its predecessor and the indicated class-$1$ fibre partner as surviving neighbours. After this prefix, changing the $a$-label from $01$ to $10$ requires at least two more $a$-coordinate changes. It also requires returning to class $0$ and subsequently entering class $1$ again. Every path therefore uses at least three $a$-coordinate changes, three class changes, and $t$ $b$-coordinate changes, for length at least $t+6$. This length is attained by following the forced prefix, changing $b$ to $1^t$ in the clean class-$1$ fibre $C^1_{01}$, returning to the clean class-$0$ fibre $C^0_{1^t}$, changing $a$ along $01,11,10$, and entering $v$.

\paragraph{The exceptional lower bound when $t=2$.}
In $EH(2,2)$, delete
\begin{equation}
 \{(10,00,0),(11,00,0)\},\qquad
 \{(00,10,0),(01,10,0)\},
\end{equation}
and take $u=(00,00,0)$ and $v=(11,10,0)$. The surviving $a$-labels of the endpoint class-$0$ fibres are respectively $\{00,01\}$ and $\{10,11\}$. With only two class changes, a path would have a single class-$1$ fibre segment, preserving $a$, and would require an $a$-label surviving in both class-$0$ fibres. No such label exists, so at least four class changes are required. At least two $a$-coordinate changes are also required. Furthermore, no surviving class-$0$ edge in either endpoint class-$0$ fibre changes the first $a$-coordinate. A path must therefore visit a class-$0$ fibre with $b\notin\{00,10\}$. A $Q_2$ walk from $00$ to $10$ through another label has length at least three, requiring at least three $b$-coordinate changes. The total length is at least $4+2+3=9$. The following surviving path has exactly that length:
\begin{equation}
\begin{aligned}
 (00,00,0)&\to(00,00,1)\to(00,01,1)\to(00,01,0)\\
 &\to(10,01,0)\to(11,01,0)\to(11,01,1)\\
 &\to(11,11,1)\to(11,10,1)\to(11,10,0).
\end{aligned}
\end{equation}
Both lower-bound families are admissible edge-structure families, and hence also admissible substructure families. They match the proved upper bound and establish both equalities.
\end{proof}
\section{The full classification and the classical comparison}

We combine the preceding cases to prove Theorem~\ref{thm:main-exact}.
\begin{proof}
Corollary~\ref{cor:exact-large} covers $3\le s\le t$. Theorem~\ref{thm:s1-exact} covers $s=1$, including the distinct structure and substructure values for $C_8$. Theorem~\ref{thm:s2-exact} covers $s=2$ and its exceptional value at $t=2$. These cases partition all positive integers $s\le t$ and prove Theorem~\ref{thm:main-exact}.
\end{proof}

\begin{corollary}\label{cor:single-class-attainment}
For every $2\le s\le t$, the full structure and substructure fault diameters are attained by $s$ parallel fault edges contained entirely in class $0$.
\end{corollary}
\begin{proof}
For $t\ge3$, the matching in Theorem~\ref{thm:general-lower} consists of edges that all change coordinate $j$ and gives the exact diameter $s+t+4$ from Theorem~\ref{thm:main-exact}. When $s=t=2$, both edges in the exceptional lower-bound construction of Theorem~\ref{thm:s2-exact} change the second $a$-coordinate and attain distance nine.
\end{proof}

For $3\le s\le t$, the ordinary diameter, classical $s$-vertex-fault diameter~\cite{Tsai2014}, and the parameters determined here satisfy
\begin{equation}
 \begin{aligned}
 D(EH(s,t))&=s+t+2,\\
 D_s^{\mathrm{vertex}}(EH(s,t))&=s+t+3,\\
 D_f(EH(s,t);K_2)=D_f^{\mathrm{sub}}(EH(s,t);K_2)&=s+t+4.
 \end{aligned}
 \label{eq:classical-comparison}
\end{equation}
Thus doubling the number of deleted vertices, from $s$ to $2s$, and exceeding the vertex connectivity $s+1$~\cite{Ma2010}, costs only one more unit of distance, provided the deleted vertices come in vertex-disjoint adjacent pairs.

\begin{remark}[One vertex fault when $s=1$]\label{rem:s1-vertex}
For every $t\ge2$, the one-vertex-fault diameter of $EH(1,t)$ is $t+5$. In the notation of Theorem~\ref{thm:s1-exact}, delete only $A_0$ and take the pair $X_0,A_{1^t}$. Every path must start with $X_0Y_0B_0$, traverse a connecting path of length three to reach the $A$-fibre, and change all $t$ coordinates of the $b$-label. Its length is at least $t+5$. The upper bound in Theorem~\ref{thm:s1-exact} includes singleton faults, so this value is exact.
\end{remark}

\section{Conclusion}

The upper-bound method has two ingredients: every surviving vertex is within two steps of a clean fibre of the other class, and a blocked cheap entrance is witnessed by a failed vertex, with witnesses for different entrances pairwise nonadjacent. Both depend on each $K_1$ or $K_2$ structure meeting at most one fibre of each class and containing at most one vertex of an independent set.

For path structures $P_k$ with $k\ge3$ and star structures $K_{1,r}$ with $r\ge2$, the connectivity thresholds are known in large parameter ranges~\cite{Liu2024,Immanuel2023,Pan2026}, but the corresponding diameters are open. Such a structure can contain several independent vertices and meet several fibres of one class, so the witness counting above no longer applies directly. A related question is how many internally vertex-disjoint short paths survive the removal of a structure-fault family, which would extend the wide-diameter results of~\cite{Tsai2014} to structure faults.

\bibliographystyle{fundam}
\bibliography{references}

\begin{thebibliography}{10}
\providecommand{\url}[1]{\texttt{#1}}
\providecommand{\urlprefix}{URL }
\expandafter\ifx\csname urlstyle\endcsname\relax
  \providecommand{\doi}[1]{doi:\discretionary{}{}{}#1}\else
  \providecommand{\doi}{doi:\discretionary{}{}{}\begingroup
  \urlstyle{rm}\Url}\fi
\providecommand{\eprint}[2][]{\url{#2}}

\bibitem{Loh2005}
Loh PKK, Hsu WJ, Pan Y.
\newblock The exchanged hypercube.
\newblock \emph{IEEE Transactions on Parallel and Distributed Systems}, 2005.
\newblock \textbf{16}(9):866--874.
\newblock \doi{10.1109/TPDS.2005.113}.

\bibitem{Tsai2014}
Tsai TH, Chen YC, Tan JJM.
\newblock Topological properties on the wide and fault diameters of exchanged
  hypercubes.
\newblock \emph{IEEE Transactions on Parallel and Distributed Systems}, 2014.
\newblock \textbf{25}(12):3317--3327.
\newblock \doi{10.1109/TPDS.2014.2307853}.

\bibitem{Krishnamoorthy1987}
Krishnamoorthy MS, Krishnamurthy B.
\newblock Fault diameter of interconnection networks.
\newblock \emph{Computers \& Mathematics with Applications}, 1987.
\newblock \textbf{13}(5--6):577--582.
\newblock \doi{10.1016/0898-1221(87)90085-X}.

\bibitem{Ma2010}
Ma M.
\newblock The connectivity of exchanged hypercubes.
\newblock \emph{Discrete Mathematics, Algorithms and Applications}, 2010.
\newblock \textbf{2}(2):213--220.
\newblock \doi{10.1142/S1793830910000590}.

\bibitem{MaZhu2011}
Ma M, Zhu L.
\newblock The super connectivity of exchanged hypercubes.
\newblock \emph{Information Processing Letters}, 2011.
\newblock \textbf{111}(8):360--364.
\newblock \doi{10.1016/j.ipl.2011.01.006}.

\bibitem{LiXu2013}
Li XJ, Xu JM.
\newblock Generalized measures of fault tolerance in exchanged hypercubes.
\newblock \emph{Information Processing Letters}, 2013.
\newblock \textbf{113}(14--16):533--537.
\newblock \doi{10.1016/j.ipl.2013.04.007}.

\bibitem{Lin2016}
Lin CK, Zhang L, Fan J, Wang D.
\newblock Structure connectivity and substructure connectivity of hypercubes.
\newblock \emph{Theoretical Computer Science}, 2016.
\newblock \textbf{634}:97--107.
\newblock \doi{10.1016/j.tcs.2016.04.014}.

\bibitem{Zhao2026}
Zhao H, Sabir E, Lin CK.
\newblock Structure fault diameter of hypercubes.
\newblock \emph{Fundamenta Informaticae}, 2026.
\newblock \textbf{196}(1):1.
\newblock \doi{10.46298/fi.14941}.

\bibitem{ZhaoSabir2026}
Zhao H, Sabir E.
\newblock On the diameter of hypercubes with star structure faults.
\newblock \emph{Discrete Applied Mathematics}, 2026.
\newblock \textbf{380}:261--272.
\newblock \doi{10.1016/j.dam.2025.10.022}.

\bibitem{Liu2024}
Liu H, Cheng D.
\newblock Structure fault tolerance of exchanged hypercube.
\newblock \emph{The Computer Journal}, 2024.
\newblock \textbf{67}(2):527--536.
\newblock \doi{10.1093/comjnl/bxac194}.

\bibitem{Immanuel2023}
Immanuel P, Greeni AB.
\newblock Characterization of minimum structure and sub-structure cut of
  exchanged hypercube.
\newblock In: Proceedings of the Ninth International Conference on Mathematics
  and Computing, volume 697 of \emph{Lecture Notes in Networks and Systems},
  pp. 249--259. Springer, 2023.
\newblock \doi{10.1007/978-981-99-3080-7_19}.

\bibitem{Pan2026}
Pan K.
\newblock Star and $h$-step-limited star connectivity of exchanged hypercubes.
\newblock \emph{Information and Computation}, 2026.
\newblock \textbf{311}:105478.
\newblock \doi{10.1016/j.ic.2026.105478}.

\end{thebibliography}
\end{document}